\documentclass[10pt,conference]{IEEEtran}
\IEEEoverridecommandlockouts
\usepackage{cite}
\usepackage{amsmath,amssymb,amsfonts}
\usepackage{algorithmic}
\usepackage{graphicx}
\usepackage{textcomp}
\usepackage{xcolor}
\usepackage{url}
\usepackage[left=1.62cm,right=1.62cm,top=1.9cm,bottom=4.44cm]{geometry}

\def\BibTeX{{\rm B\kern-.05em{\sc i\kern-.025em b}\kern-.08em
    T\kern-.1667em\lower.7ex\hbox{E}\kern-.125emX}}

\usepackage{pbalance}

\makeatletter
\newcommand{\linebreakand}{%
  \end{@IEEEauthorhalign}
  \hfill\mbox{}\par
  \mbox{}\hfill\begin{@IEEEauthorhalign}
}
\makeatother
\begin{document}
\title{Improving the Security of Containerized Workloads using Transparency and Traceability Services}

\author{%
\IEEEauthorblockN{
Nikos Fotiou,\IEEEauthorrefmark{1}
Lefteris Georgiadis,\IEEEauthorrefmark{1}
Ignacio Lacalle, \IEEEauthorrefmark{2}
George C.  Polyzos,\IEEEauthorrefmark{1}\IEEEauthorrefmark{3}\IEEEauthorrefmark{4}
Vasilios A. Siris\IEEEauthorrefmark{1}\IEEEauthorrefmark{3}
}
\IEEEauthorblockA{\IEEEauthorrefmark{1}
ExcID, Athens, Greece}
\IEEEauthorblockA{\IEEEauthorrefmark{2}
Communications Department, Universitat Politècnica de València, Valencia, Spain}
\IEEEauthorblockA{\IEEEauthorrefmark{3}
Mobile Multimedia Laboratory, Department of Informatics,\\
School of Information Sciences and Technology, Athens University of Economics and Business, Greece}
\IEEEauthorblockA{\IEEEauthorrefmark{4}
School of Data Science,
Chinese University of Hong Kong,
Shenzhen, China}
}

\maketitle
\begin{abstract}

Containerized workloads are commonly built via CI/CD pipelines, stored in registries, and executed across heterogeneous infrastructures, including cloud and edge environments. A single compromised build step or credential can turn routine automation into large-scale distribution of malicious artifacts, motivating integrity, transparency, and enforceable deployment-time checks. In this paper, we present an architecture for verifiable container image distribution that addresses key-management challenges and enables policy-enforced admission-time verification. A transparency service generates one-time signing keys bound to authenticated identities, records signing events in an append-only transparency registry, and returns cryptographically verifiable proofs of inclusion. These proofs and identity attributes are attached to image metadata and evaluated by policy-as-code at admission time, so only compliant artifacts are deployed. We implement a proof-of-concept integrated with GitHub Actions and GitLab Runners and evaluate how the resulting pipeline mitigates common supply-chain attacks under a realistic threat model.

% The abstract should briefly summarize the contents of the paper in
% 150--250 words.

\end{abstract}
\section{Introduction}
\label{sec:introduction}

Containerization is a dominant packaging and delivery mechanism for modern software: images are built in CI/CD pipelines, stored in registries, and deployed via orchestrators across cloud, enterprise, and edge infrastructures. This automation improves delivery speed and scale, but it also expands the software supply-chain attack surface~\cite{Pie2023,Ben2022}. A common taxonomy distinguishes attacks that target (i) upstream dependencies, (ii) build and release infrastructure, and (iii) developers~\cite{Will2025}. In this work, we focus on the second vector, which is particularly relevant to CI/CD-driven container image distribution.

Attacks on build infrastructure can be mitigated through a combination of \emph{integrity} mechanisms (cryptographic signing) and \emph{transparency} mechanisms (tamper-evident logging and independent audit)~\cite{Will2025}. In practice, however, signing adoption and key management remain challenging~\cite{Sch2025}, and deployment-time verification is often inconsistently enforced~\cite{Kal2025}. Moreover, many approaches place signing capability within the CI/CD environment, so a compromise of runner credentials or build systems can enable attackers to produce seemingly valid signatures for malicious artifacts.

\textbf{Our contribution:} We present an end-to-end approach for verifiable container image distribution that combines identity-bound signing, transparency logging, and policy-enforced admission. Concretely, we (i) introduce a transparency service that generates one-time signing keys, issues short-lived, identity-bound certificates, and records each signing event in an append-only transparency registry, (ii) extend image metadata with traceability evidence (notably proofs of inclusion) so that deployments can validate provenance independently of the image registry, and (iii) express deployment-time checks as policy-as-code that verifies signatures, authorized signer identity, and transparency evidence before execution. We implement a proof-of-concept integrated with GitHub Actions and GitLab Runners. By avoiding reusable signing keys in the CI/CD environment and making signing events auditable, the approach reduces the impact of CI/CD compromise and enables detection of signing misuse via transparency monitoring.

The remainder of this paper is organized as follows: Section~\ref{sec:background} provides background. Section~\ref{sec:system-design} presents the design. Section~\ref{sec:implementation} describes implementation and evaluation. Section~\ref{sec:related} reviews related work, and Section~\ref{sec:conclusion} concludes our paper.

\section{Background}
\label{sec:background}
\subsection{DevSecOps for Containerized Workloads}

DevSecOps~\cite{Raj2022} integrates security into CI/CD by running automated checks (e.g., SAST, SCA, secret scanning, and image scanning) and enforcing compliance as code throughout build and deployment. For containerized workloads this typically culminates in publishing signed images and associated metadata to registries and deploying them via automated reconciliation. Heterogeneous deployments---including cloud--edge settings---amplify the need for reliable, low-touch authenticity and provenance verification at admission time.
\subsection{Transparency Registries}

Transparency registries are append-only, publicly verifiable logs that provide auditability in software supply chains by recording artifact creation and signing events. A verifier can check that an event is logged via a \emph{proof of inclusion} derived from a Merkle-tree commitment~\cite{hu2021}, making unauthorized substitutions detectable. Certificate Transparency (CT) is a widely deployed instance of this model~\cite{rfc6962,rfc9162}.

In our context, the registry logs container image signing events (e.g., digest, signer identity, and related metadata). Deployment sites retrieve and verify the corresponding inclusion proof before admitting an artifact, enabling independent provenance checks even when the registry or transport path is untrusted.

\section{System Design}
\label{sec:system-design}

\subsection{Overview}
\begin{figure}
\begin{centering}
\includegraphics[width=0.8\linewidth]{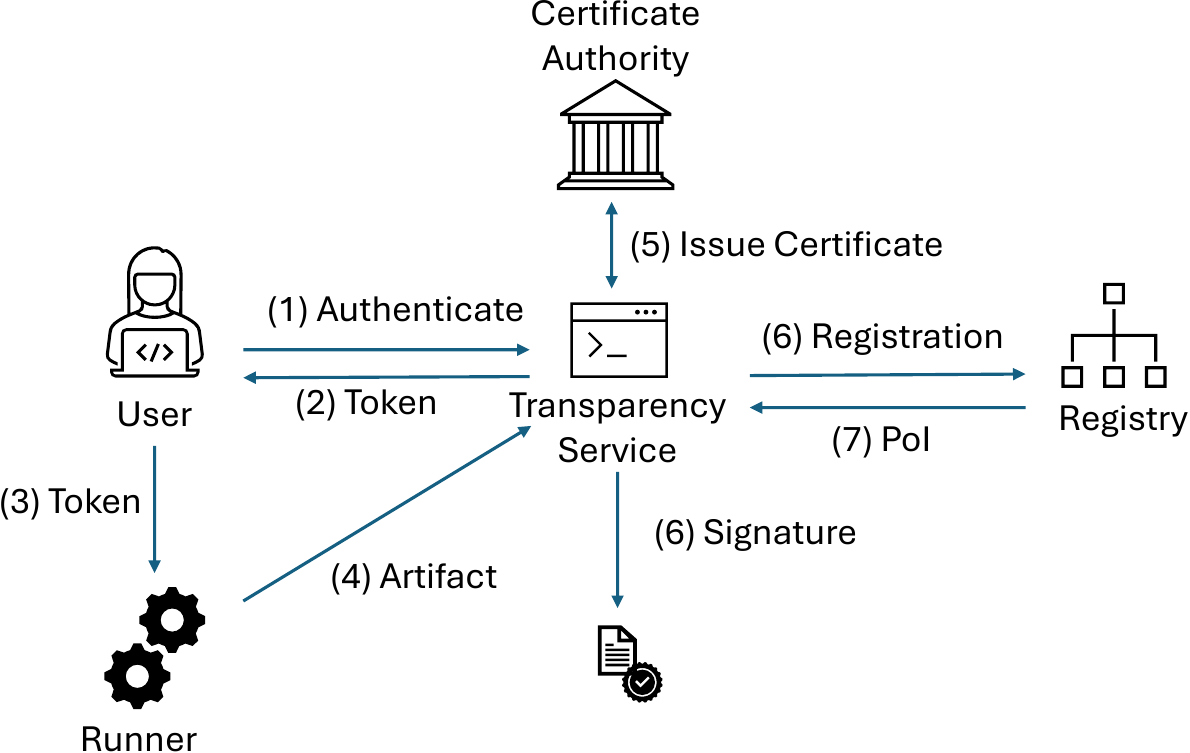}
\caption{Components of our solution.}
\label{fig-overview}
\end{centering}
\end{figure}

At the core of our approach lies a \textit{transparency service} responsible for both signing artifacts and recording signing events in a transparency registry (see also Fig.~\ref{fig-overview}). The transparency registry is provided as public service by a trusted third party. The transparency service is integrated into the CI/CD pipeline, enabling automated, verifiable signing of container images during the build process. A transparency service may be administrated by an \emph{organization} or provided as a service by a trusted third party. 

\emph{Users} of an organization authenticate to the transparency service and obtain access tokens bound to their organization-specific identity. These tokens are then configured in the corresponding CI/CD ``runners'' of the source code management system. Whenever a pipeline stage invoking the transparency service is executed by such a runner, the service generates an one-time signing key. The corresponding public key is embedded in a short-lived certificate that also includes the authenticated user’s identity. Using this certificate, the service signs the provided artifact, in our case, a container image digest or provenance metadata. Once the signature is produced, both the signature and the signing certificate are submitted to the transparency registry. The registry records the signing event in its append-only Merkle-tree-based log and returns a \textit{proof of inclusion} (PoI), which cryptographically attests that the event is logged and immutable. The proof of inclusion, along with the signature and certificate, are embedded into a manifest file associated with the container image. This manifest is stored in the container registry alongside the image itself.

On the deployment side, a component continuously observes the container registry for newly published images. Upon detecting an update, it retrieves the image and its associated manifest. Before deployment, it verifies the image signature against the included certificate, validates the proof of inclusion against the transparency registry’s Merkle root, and ensures that the certificate’s identity corresponds to an authorized user as defined by a local policy. Only if all verification steps succeed is the container image admitted for execution. This ensures that all deployed containers originate from trusted build processes, are signed by authorized users, and are cryptographically traceable via the transparency registry.

\subsection{Transparency registry monitoring}

An integral component of transparency registries is the presence of \emph{monitors}. These entities continuously observe the registry for new events, ensuring that all log entries are visible and auditable shortly after their creation. Organizations may establish a trust relationship with an external monitor, or operate their own internal monitoring infrastructure, to receive timely notifications about events related to their artifacts.  

When properly configured, a monitor can detect and report a range of events of interest. A benign example is the legitimate signing of a new container image by an authorized user within the organization. In contrast, a more severe case could involve the detection of a suspicious signing event, such as one resulting from a compromised organizational signing key or unauthorized use of build infrastructure. By correlating these events with internal records and policies, organizations can determine whether the activity was authorized.  

Upon detecting a potentially malicious event, the organization should initiate an appropriate incident response process. This may involve revoking the compromised key, halting deployments of the affected container image, or issuing alerts to downstream consumers.

\subsection{Trust Model}
\label{sec:trust-model}

We now detail the trust model of our solution. Let $\mathcal{O}$ denote an organization, $\mathcal{U}$ the set of its users, and let
\[
\textsf{CA},\ \textsf{TS},\ \textsf{TR},\ \textsf{CR},\ \textsf{IdP},\ \textsf{D}
\]
denote, respectively, a Certificate Authority, the Transparency Service, the Transparency Registry, a (container) image registry, the organization’s OpenID Connect Identity Provider, and a deployment site (cloud or edge cluster). We write $u \in \mathcal{U}$ for a user identity belonging to $\mathcal{O}$.

\paragraph{Setup and long-term trust anchors.}
During setup, the CA generates a key pair $(sk_{\mathsf{CA}}, pk_{\mathsf{CA}})$ used exclusively for signing short-lived end-entity certificates. The public key $pk_{\mathsf{CA}}$ is provisioned as a trust anchor at deployment site $\textsf{D}$. The registry $\textsf{TR}$ exposes an append-only, Merkle-tree-based log with publicly verifiable inclusion proofs and supports consistency auditing by independent monitors. We assume an adequate ecosystem of external auditors exists; the design and operation of such auditing is out of scope of this paper.

\paragraph{Identity and authentication.}
Users authenticate to the Transparency Service $\textsf{TS}$ through the organization’s Identity Provider $\textsf{IdP}$ using OpenID Connect.

\paragraph{Ephemeral signing credentials.}
For each signing event $s$, the TS generates an ephemeral key pair $(sk_s, pk_s)$ and issues a short-lived certificate

\[
C_s \leftarrow \mathrm{Certify}_{sk_{\mathsf{CA}}}\!\big(pk_s,\ \mathrm{id}(u),\ [t_s^{\mathrm{start}}, t_s^{\mathrm{end}}]\big),
\]
binding the public key $pk_s$ to the authenticated identity $\mathrm{id}(u)$ with a validity window $[t_s^{\mathrm{start}}, t_s^{\mathrm{end}}]$. The TS then signs the artifact digest $d$ to produce
\[
\sigma_s \leftarrow \mathrm{Sign}_{sk_s}(d).
\]
\textbf{Key-destruction assumption:} $\textsf{TS}$ securely deletes $sk_s$ immediately after use; only $C_s$ and $\sigma_s$ persist. 

\paragraph{Transparency logging and time.}
The TS submits $(\sigma_s, C_s)$ to the transparency registry $\textsf{TR}$, which returns a proof of inclusion
\[
\mathrm{PoI}_s = (\pi_s,\ R_t,\ t^{\mathrm{log}}_s),
\]
where $\pi_s$ is the Merkle authentication path, $R_t$ the current log root commitment, and $t^{\mathrm{log}}_s$ the registry’s timestamp for the leaf corresponding to the event. The triplet $(\sigma_s, C_s, \mathrm{PoI}_s)$ is embedded into the image’s manifest and stored alongside the image at the container registry $\textsf{CR}$.

\paragraph{Policy and authorization.}
Each deployment site $\textsf{D}$ maintains a \emph{policy} $\mathcal{P}$ that governs which identities and artifacts are admissible. At minimum, $\mathcal{P}$ specifies: (i) the trusted CA key $pk_{\mathsf{CA}}$; (ii) the set of authorized signer identities $\mathcal{A} \subseteq \{\mathrm{id}(u) : u \in \mathcal{U}\}$; and (iii) freshness and transparency requirements. Policies may additionally encode constraints on algorithms, key sizes, permitted issuers, SBOM presence, SLSA provenance attributes, vulnerability thresholds, and acceptable registries, but we focus here on the core trust predicates.

\paragraph{Acceptance predicate.}
Let $m$ denote a retrieved manifest containing $(d, \sigma_s, C_s, \mathrm{PoI}_s)$. The deployment site applies a deterministic acceptance predicate
\[
\mathrm{Accept}_{\mathcal{P}}(m,\textsf{TR}) \in \{\textsf{true},\textsf{false}\}
\]
defined as the conjunction of four verifications:
\begin{align}
\textbf{Sig:} &\quad \mathrm{Verify}_{pk(C_s)}(\sigma_s, d) = \textsf{true}, \label{eq:sig}\\
\textbf{AuthZ:} &\quad \mathrm{id}(C_s) \in \mathcal{A} \ \wedge\ \mathrm{VerifyCert}_{pk_{\mathsf{CA}}}(C_s) = \textsf{true}, \label{eq:authz}\\
\textbf{Transp:} &\quad \mathrm{VerifyPoI}\!\big(PoI\big) = \textsf{true}, \label{eq:poi}\\
\textbf{Fresh:} &\quad t^{\mathrm{log}}_s \in \big[t_s^{\mathrm{start}},\ t_s^{\mathrm{end}}\big]. \label{eq:fresh}
\end{align}
Here, $pk(C_s)$ and $\mathrm{id}(C_s)$ extract the public key and identity from the certificate, $\mathrm{VerifyCert}_{pk_{\mathsf{CA}}}$ validates the certificate chain and validity interval, and $\mathrm{VerifyPoI}_{\textsf{TR}}$ verifies $\mathrm{PoI}_s$.

The manifest is \emph{admitted} for deployment if and only if $\mathrm{Accept}_{\mathcal{P}}(m,\textsf{TR})=\textsf{true}$. Equation~\eqref{eq:sig} ensures integrity of the digest under the ephemeral key bound to the recorded identity; Equation~\eqref{eq:authz} enforces authorization against organizational policy; Equation~\eqref{eq:poi} provides append-only transparency and auditability; and Equation~\eqref{eq:fresh} ties the log event to the certificate’s validity.

\paragraph{Trust assumptions.}
Our model relies on the following assumptions:
\begin{enumerate}
  \item \textbf{CA correctness and key protection.} The CA issues certificates only for the considered Transparency Service (TS) and protects $sk_{\mathsf{CA}}$ (e.g., via HSM). 
  \item \textbf{TS soundness and ephemerality.} The TS correctly authenticates users via $\textsf{IdP}$, enforces identity uniqueness, generates fresh ephemeral keys per signing event, and destroys $sk_s$ immediately after use.
  \item \textbf{Transparency Registry append-only property.} The transparency registry $\textsf{TR}$ implements an append-only Merkle log with publicly verifiable inclusion proofs and is subject to sufficient external monitoring to detect equivocation or rollback.
  \item \textbf{Container image registry security.} Deployment sites can securely access the correct container image registry.
\end{enumerate}

\section{Implementation}
\label{sec:implementation}

We have implemented a transparency service that can be deployed as a Docker container\footnote{\url{https://github.com/excid-io/staas}}. The service issues temporary signing certificates using a locally deployed instance of the Fulcio Certificate Authority (CA)\footnote{\url{https://docs.sigstore.dev/certificate_authority/overview/}}, configured to use a certificate signing key stored in a Hardware Security Module (HSM) for secure key management. This setup ensures that signing operations are performed using short-lived credentials, mitigating the risk of key compromise.

Our transparency service has been integrated with both GitHub Actions and GitLab Runners. Whenever a container image is built, the corresponding CI/CD runner submits a signing request to the service. Each request includes the container image digest, a Software Bill of Materials (SBOM), and a Supply Chain Levels for Software Artifacts (SLSA) provenance artifact. The service signs these artifacts using the temporary certificate and records the signing event in the Rekor transparency registry\footnote{\url{https://docs.sigstore.dev/logging/overview/}}, a publicly accessible append-only log provided as a public good by the Open Source Security Foundation (OSSF). A registry monitoring service has been implemented using Rekor monitor\footnote{https://github.com/sigstore/rekor-monitor}.
Log entries typically appear in the Rekor transparency log approximately 12 minutes after submission.

To automate deployment, the container registry is continuously monitored using FluxCD\footnote{\url{https://fluxcd.io}}, which detects the availability of new container images. Upon detecting a new image, FluxCD pulls it into the local cluster, where it is subjected to a policy-based verification process. This process is enforced by Kyverno\footnote{\url{https://kyverno.io}}, a cloud-native policy engine for Kubernetes. Kyverno verifies that the container’s manifest includes (i) a certificate issued to a valid organizational user, containing a public key that can be used to verify the container signature, and (ii) a valid proof of inclusion from the transparency registry. Only if both conditions are satisfied is the image admitted into the cluster for deployment. Since all operations are based on the evaluation of traditional signatures, the overhead of our solution is negligible. 

\subsection{Security Evaluation}
\label{sec:security-eval}

We evaluate the security of our implementation and show how the verification and policy mechanisms described in Sections~\ref{sec:system-design} mitigate concrete software supply-chain attacks against container image distribution and deployment workflows, including those encountered in cloud--edge deployments.

\paragraph{Threat model.}
We assume the following types of adversary $\mathcal{A}$:
(i) a \emph{network attacker} who can intercept, delay, drop, or modify traffic between the container registry \textsf{CR} and a deployment site \textsf{D};
(ii) a \emph{registry adversary} who can upload artifacts to \textsf{CR} (e.g., via a compromised publisher account or a misconfigured registry) but has no control over the cluster admission at \textsf{D};
(iii) a \emph{credential adversary} who steals or misuses signing-related credentials (e.g., obtains a runner token or compromises a developer account) to induce the Transparency Service \textsf{TS} to produce a valid signature and short-lived certificate for a malicious image digest.
We assume the Certificate Authority \textsf{CA} key $sk_{\mathsf{CA}}$ is protected by an HSM and not compromised; the Transparency Registry \textsf{TR} provides append-only Merkle logging with publicly verifiable inclusion proofs and is monitored by independent auditors; and clocks for \textsf{TS} and \textsf{TR} are loosely synchronized with skew bound $\Delta$ (Section~\ref{sec:trust-model}). As in Section~\ref{sec:trust-model}, each manifest $m$ carries $(d,\sigma_s,C_s,\mathrm{PoI}_s)$ and is admitted only if the acceptance predicate $\mathrm{Accept}_{\mathcal{P}}(m,\textsf{TR})$ holds, i.e., the conjunction of~\eqref{eq:sig}--\eqref{eq:fresh}.

\subsubsection*{Attack 1: Image modification in transit}
\textbf{Goal.} Modify a container image or its manifest on the path from \textsf{CR} to \textsf{D} so that a tampered artifact is deployed.

\textbf{Analysis.} Any bit-level modification to the image blob changes its digest $d$. Since $\sigma_s = \mathrm{Sign}_{sk_s}(d)$ is bound to $d$, verification~\eqref{eq:sig} fails for a tampered image. Likewise, swapping the manifest while keeping the image fixed causes a digest–signature mismatch detectable by~\eqref{eq:sig}. An adversary could attempt \emph{manifest swapping} with a consistent tuple $(d',\sigma'_s,C'_s,\mathrm{PoI}'_s)$ for \emph{another} image; in that case, integrity check~\eqref{eq:sig} may pass for $d'$, but the deployment will use the wrong image (a classic \emph{mix-and-match} risk). Our pipeline prevents this because (a) the admission controller at \textsf{D} computes the digest of the fetched image and compares it against $d$ from $m$ prior to scheduling; and (b) the PoI is verified for the canonical leaf value derived from $(d,\sigma_s,C_s)$ (Equation~\eqref{eq:poi}), binding the transparency evidence to the exact digest to be deployed. Therefore, any in-transit modification or digest/manifest mismatch is rejected.

\subsubsection*{Attack 2: Unauthorized image in the registry}
\textbf{Goal.} Place an unapproved (possibly unsigned) image in \textsf{CR} and get it deployed.

\textbf{Analysis.} Unauthorized uploads may occur via compromised publisher accounts or misconfiguration in \textsf{CR}. Our defense relies on \emph{policy enforcement}: even if an attacker places an image in \textsf{CR} image admission will fail. Particularly, if the image is \emph{unsigned} or signed with an untrusted CA or signed by an unauthorized user \eqref{eq:sig} or \eqref{eq:authz} fails.

\subsubsection*{Attack 3: Key or identity breach (misuse of signing capability)}
\textbf{Goal.} Use stolen credentials or compromised CI runner secrets to induce \textsf{TS} to sign a malicious digest $d^\star$ and obtain $(\sigma^\star, C^\star, \mathrm{PoI}^\star)$ that passes admission.

\textbf{Analysis.} Our design promotes \emph{ephemerality} and \emph{traceability}. Each signing event uses a freshly generated key $sk_s$ destroyed immediately after signing, and each certificate $C_s$ has a short validity window $[t_s^{\mathrm{start}}, t_s^{\mathrm{end}}]$. If $\mathcal{A}$ reflects the organization’s authorized identities, the attack is \emph{prevented} when the stolen identity is not in $\mathcal{A}$ (Equation~\eqref{eq:authz}). If, however, the attacker compromises an \emph{authorized} identity, admission may \emph{initially} succeed because all checks in~\eqref{eq:sig}–\eqref{eq:fresh} hold. In this case, \emph{detection} is provided by transparency monitoring: $\mathrm{PoI}^\star$ makes the malicious signing event publicly visible and linkable to $\mathrm{id}(C^\star)$ and $d^\star$. Let $T_{\mathrm{mon}}$ denote the configured monitor interval. Then the \emph{detection delay} $D$ satisfies
\[
0 \le D \le T_{\mathrm{mon}}
\]
under a uniform event-arrival assumption. Thus, detection efficiency scales with the monitoring cadence: between two monitoring events, misuse may go undetected.

\textbf{Mitigation and response.} Upon detection, the organization must execute an incident response plan, e.g., (i) immediately \emph{remove} the compromised identity from $\mathcal{A}$ and rotate runner tokens; (ii) \emph{deny-list} the malicious digest(s) $d^\star$ at admission, and update GitOps state to block further rollout; (iii) \emph{quarantine or rollback} any deployed instances; (iv) \emph{rebuild and reissue} corrected images with fresh signing events; and (v) \emph{purge} or deprecate the malicious images from \textsf{CR} (subject to registry capabilities and retention rules). Note that \textsf{TR} is append-only; the malicious event cannot be erased, but recording a revocation/incident attestation provides durable evidence and supports downstream denylists and monitors.

\paragraph{Security properties.}
Relative to the acceptance predicate in Section~\ref{sec:trust-model}, our implementation enforces the following:

\noindent\textbf{Authenticity and integrity.}
If $\mathrm{Verify}_{pk(C_s)}(\sigma_s,d)=\textsf{true}$ and $\mathrm{VerifyCert}_{pk_{\mathsf{CA}}}(C_s)=\textsf{true}$, then the image digest $d$ was signed by the holder of the ephemeral private key certified by \textsf{CA} for identity $\mathrm{id}(C_s)$ within the certificate’s validity interval (barring CA compromise). Any modification to the image or mismatch between image and manifest causes~\eqref{eq:sig} to fail.

\noindent\textbf{Provenance (transparency binding).}
If $\mathrm{VerifyPoI}_{\textsf{TR}}(d,\sigma_s,C_s,\pi_s,R_t)=\textsf{true}$, then the tuple $(d,\sigma_s,C_s)$ is committed in \textsf{TR}’s append-only log. Under standard Merkle-tree assumptions and monitor coverage, equivocation or deletion is detectable. Freshness~\eqref{eq:fresh} prevents replay of stale but logged signatures.

\noindent\textbf{Authorization.}
The combination of identity-bound certificates and  ensures that only designated principals can produce admissible images. Policy updates propagate at the next reconcile, preventing further admission.

\paragraph{Discussion}
An important mitigation against key or identity breaches is the \emph{detection} of offending signing events via the transparency registry, but detection latency depends on the monitoring frequency. Let $T_{\mathrm{mon}}$ be the monitoring interval; as discussed, the detection delay $D$ satisfies $0 \le D \le T_{\mathrm{mon}}$. To bound the risk of prematurely admitting a malicious image that has just been logged, a deployment site can enforce an \emph{acceptance delay} (or cooling-off window) $T_{\mathrm{acc}}$: a manifest is considered admissible only after its corresponding transparency event has persisted in the registry for at least $T_{\mathrm{acc}}$. This configuration creates an explicit trade-off between rapid vulnerability remediation (small $T_{\mathrm{acc}}$) and breach-mitigation efficacy (large $T_{\mathrm{acc}}$); organizations should tune $T_{\mathrm{mon}}$ and $T_{\mathrm{acc}}$ according to risk appetite, criticality of the workloads, and expected incident response times. 

A second operational consideration is \emph{inclusion latency} at transparency registries. Merkle-tree-based logs typically batch submissions and integrate them periodically, so a submission initially receives a \emph{promise of inclusion} (a signed receipt acknowledging intent to include) and only later yields a verifiable proof of inclusion (PoI) once the leaf is incorporated into a committed tree root. In practice, inclusion latency can range from minutes (e.g., single-digit to tens of minutes in software supply-chain logs) to many hours (e.g., up to 24~hours in some WebPKI CT logs). Our admission logic should therefore distinguish between \emph{promise} and \emph{proof}: deployment sites may tentatively stage artifacts upon receipt of a promise but should admit them only after verifying that the promise has been \emph{fulfilled} by an actual PoI consistent with the registry’s latest checkpoint. Unlike the WebPKI browser model—where online PoI checks are often avoided for privacy reasons—our setting does not involve user browsing behavior; thus, deployment sites can safely perform online queries to confirm that a promised entry has been included before admission.

\section{Related Work}
\label{sec:related}
Our transparency service closely resembles the functionality provided by Cosign~\cite{sigstore}, but introduces several key improvements. First, whereas Cosign relies on a public instance of the Fulcio CA, which is pre-configured to issue certificates only for a restricted set of Identity Providers, our approach is agnostic to the choice of Identity Provider. This flexibility enables organizations to integrate their own IdPs (e.g., enterprise-specific OIDC providers) without being constrained by the defaults of a public service. Second, in a Cosign-based deployment, signing keys are generated locally by the CI/CD infrastructure. By contrast, in our design all ephemeral keys are generated centrally by the transparency service. This service can be isolated and hardened, e.g., deployed within a protected enclave or supported by hardware security modules—thus reducing the number of attack vectors and improving the overall trustworthiness of the signing workflow by reducing the attack surface. Furthermore, in our solution a compromised CI/CD pipeline can denied access to the transparency service, therefore our system is also resilient against this types of attack. 

OpenPubKey~\cite{Hei2023} provides an alternative mechanism for binding cryptographic keys to user identities by leveraging the OpenID Connect protocol. In this approach, a hash of the user’s public key is embedded as a digest within the issued ID token, which can then be used in place of a traditional certificate. This design effectively eliminates the need for a dedicated Certificate Authority. However, replacing certificates with ID tokens introduces compatibility challenges, as it requires modifications to existing signature verification software that is typically built around X.509 certificate chains. In addition, refreshing a signing key necessitates a new interaction with the Authorization Server to obtain an updated ID token. Nevertheless, integration with OpenPubKey is considered as a future work item.

Our work intersects with several established efforts in software supply chain security. The Update Framework (TUF)~\cite{tuf}, which provides information about software updates,  in-toto~\cite{intoto}, which provide mechanisms for resilient signed artifact distribution and supply chain provenance tracking, Software Bill of Materials (SBOM)~\cite{Gar2025} and the Supply-chain Levels for Software Artifacts (SLSA) framework~\cite{Tam2024}, which standardize the description of dependencies and build provenance. Despite these advances, existing solutions rarely integrate real-time transparency logging, ephemeral credential issuance, user-bound certificate generation, and policy-enforced verification as a unified framework. Our design combines these elements, enabling signatures with ephemeral, identity-bound certificates, logged in a transparency registry, and verified at deployment time using policy engines such as Kyverno. Therefore, our solution can be used as a signing backend for all these efforts. 

Information recorded in transparency registries is publicly accessible, which may also benefit adversaries. For instance, in the Web PKI ecosystem, Certificate Transparency logs have been exploited by attackers to identify new domains and launch targeted attacks~\cite{Kon2022}. To mitigate such risks, several research efforts have explored the notion of \emph{privacy-preserving signing} (e.g., Speranza ~\cite{Mer2023}). Integration of such privacy-preserving techniques into our framework is considered as a promising direction for future work.

\section{Conclusion}
\label{sec:conclusion}

This paper presented a verifiable approach to securing container image distribution by combining one-time, identity-bound signing with transparency logging and policy-enforced admission. By embedding proofs of inclusion and signer identity information alongside signatures, deployment sites can independently validate integrity, provenance, and freshness before execution.

We implemented a proof-of-concept integrated with GitHub Actions and GitLab Runners and demonstrated admission-time enforcement within a GitOps pipeline (e.g., Kyverno). Future work includes blockchain-backed transparency registries, systematic tuning of monitoring and acceptance trade-offs, and alternative identity-binding and privacy-preserving signing mechanisms (e.g., OpenPubKey and Speranza).

\section*{Acknowledgement} 
The work reported in this paper has been partly funded by the EU’s Digital Europe Programme project CRA Made Easy (CRACY) under grant agreement No 101190492.

%
% ---- Bibliography ----
%
% BibTeX users should specify bibliography style 'splncs04'.
% References will then be sorted and formatted in the correct style.
%
% \bibliographystyle{splncs04}
% \bibliography{mybibliography}
%
% Author, F.: Article title. Journal \textbf{2}(5), 99--110 (2016)

% \bibitem{ref_lncs1}
% Author, F., Author, S.: Title of a proceedings paper. In: Editor,
% F., Editor, S. (eds.) CONFERENCE 2016, LNCS, vol. 9999, pp. 1--13.
% Springer, Heidelberg (2016). \doi{10.10007/1234567890}

% Author, F., Author, S., Author, T.: Book title. 2nd edn. Publisher,
% Location (1999)

\bibliographystyle{IEEEtran}
\bibliography{IEEEabrv,references}

\end{document}